# The astrophysics of rotational energy extraction from a black hole

David Garofalo[1] & Chandra B. Singh[2]
1. Department of Physics, Kennesaw State University, USA, (orcid: 0000-0001-5536-829X) email:dgarofal@kennesaw.edu
2. South-Western Institute for Astronomy Research, Yunnan University, University Town, Chenggong, Kunming 650500, China, (orcid: 0000-0002-7782-5719) email: chandrasingh@ynu.edu.cn

**Abstract**

Recent work has called into question whether nature can extract the rotational energy of a black hole via electromagnetic fields by appealing to an alleged ability to absorb current. We describe the strategies needed to properly treat the astrophysics in curved spacetime near black holes, showing that while the Blandford-Znajek effect is sound, the deeper nature of the electric nature of black holes remains unresolved.

1. Introduction

Recently King & Pringle[1] have challenged the Blandford-Znajek mechanism for the extraction of black hole rotational energy by suggesting that the electric field near the horizon gets shorted out by the black hole's ability to absorb charge. That highly curved spacetime near black hole horizons leaves an indelible imprint on the accreting plasma is certain, but we show that it is the plasma astrophysics that holds the key to understanding the relevant electrical properties of the Blandford-Znajek process. Much of the confusion likely stems from the 3+1 form of the magnetohydrodynamic equations in curved spacetime, that avoid both a full covariant treatment and the familiar vector formulation. As a result, those approaching the subject from an astrophysical perspective, often acquire the false impression that black holes behave dramatically differently than other regions of empty spacetime. In this work, we try to make amends by illustrating the process needed to determine the electrical structure of the near black hole region, with emphasis on the increased complexity in more realistic astrophysical environments (see also ref.[2]). Ultimately, this leads us to the recognition that in order to determine the astrophysical behavior of electric fields and currents near black holes, we must implement the covariant and causal general relativistic Ohm's law

While the black hole horizon cannot affect the physics of the magnetosphere, it has long been recognized that the near horizon region must influence the dynamics of the inflowing plasma. This requires evaluating the equations of the accretion flow on the so-called stretched horizon[3] to produce a 'regularity' condition[4]. While the highly curved spacetime near the horizon alters the form of the equations[5], it cannot invalidate the physical content of the equations. In other words, black holes cannot determine or change the causal nature of the physics of the inflowing plasma, which, instead, is determined by the plasma itself. Accordingly, the nature of the electrical properties of the accreting flow near a black hole, is determined by the astrophysical parameters of the flow. As the plasma astrophysics complicates for realistic flows, various simplifications are adopted. In this work, we explore the strategy for determining the electrical properties near black holes for increasingly realistic flows. What we show is that while much of this remains unresolved currently, the criticisms mentioned above amount to ignoring the strategy that is needed to determine the behavior of the flow near black holes.

In Section 2.1 we explore the strategy for determining the electric field near black holes in the ideal magnetohydrodynamic limit. In Section 2.2 we extend that to a more realistic astrophysical flow by implementing a finite conductivity. We then discuss the problems with causality in that context. This will take us in Section 2.3 to generalize Ohm's law to include time dependence and particle species, showing that the appropriate condition for the relevant near horizon electric fields necessary for the Blandford-Znajek mechanism, remain unresolved. We then summarize and conclude.

## 2. Physics of the accreting fluid

### 2.1 – Ideal MHD

Ideal MHD is a cornerstone of both analytic[3] as well as numerical work[6,7] on the Blandford-Znajek mechanism. It amounts to the statement that no dissipation occurs within a perfectly conducting medium where the electric field in the fluid frame must vanish. This is captured by the following equation

$$F_{ab}U^b = 0, \qquad (1)$$

where $F_{ab}$ is the Faraday tensor and $U^b$ is the 4-velocity of the accreting fluid. The non-covariant and more familiar version of this equation is the set of equations

$$\mathbf{E} \cdot \mathbf{v} = 0 \qquad (2)$$

$$\mathbf{E} + \mathbf{v} \times \mathbf{B} = 0 \qquad (3)$$

where **E**, **v** and **B** are the electric field, the fluid velocity, and the magnetic field, respectively. Of course, equation (2) is implied by equation (3) and so is

$$\mathbf{E} \cdot \mathbf{B} = 0. \qquad (4)$$

This is a highly idealized, and ultimately problematic, set of equations. In addition to the infinite conductivity, and the absence of a current constraint in these equations (it is obtained via Ampere's law), the fluid does not distinguish between particle species. If there is any place in the universe where such an idealization is poorly motivated, it is in the extreme environment near a black hole. From a circuit perspective, the induced voltage or emf near the black hole requires integrating the induced electric field, which from (3) and up to a minus sign is

$$\varepsilon = \int \mathbf{E} \cdot d\mathbf{l} = \int -(\mathbf{v} \times \mathbf{B}) \cdot d\mathbf{l} \qquad (5)$$

where the integration involves dl with dr = 0, where r is the radial coordinate. The Boyer-Lindquist electric field components $F_{10}$, $F_{20}$, and $F_{30}$, in equation (1) are obtained by

$$F_{rb}U^b = F_{rt}U^t + F_{r\theta}U^\theta + F_{r\phi}U^\phi = 0 \qquad (6)$$
$$F_{\theta b}U^b = F_{\theta t}U^t + F_{\theta r}U^r + F_{\theta \phi}U^\phi = 0$$
$$F_{\phi b}U^b = F_{\phi t}U^t + F_{\phi r}U^r + F_{\phi \theta}U^\theta = 0 .$$

Axi-symmetry and zero radial and poloidal velocities (i.e. $U^r = U^\theta = 0$) at the horizon reduces the system to

$$F_{rt}U^t = -A_{\phi,r}U^\phi \qquad (7)$$

where we have implemented the relation between the Faraday tensor and the vector potential

$$F_{cb} = A_{b,c} - A_{c,b}. \qquad (8)$$

But the invariant magnetic flux on the black hole is

$$\phi_{BH} = \int A_3 dx^3 = \int A_\phi d\phi = 2\pi A_\phi \qquad (9)$$

Hence,

$$F_{rt}U^t = -A_{\phi,r} U^\phi = -(\partial \phi_{BH}/\partial r)U^\phi/2\pi \qquad (10)$$

which leads to

$$F_{rt} = -(\partial \phi_{BH}/\partial r)U^\phi/(2\pi U^t) . \qquad (11)$$

Now, $U^\phi/U^t = d\phi/dt$ ($\phi$ is the azimuthal Boyer-Lindquist coordinate) is the angular velocity of the fluid with respect to the Boyer-Lindquist frame in which the electric field components are determined. For the second Boyer-Lindquist electric field component we have

$$F_{\theta t}U^t = -F_{\theta r}U^r - F_{\theta \phi}U^\phi = -A_{\phi,\theta}U^\phi = -(\partial \phi_{BH}/\partial \theta)U^\phi/2\pi \qquad (12)$$

which gives

$$F_{\theta t} = -(1/2\pi)(\partial \phi_{BH}/\partial \theta) d\phi/dt. \qquad (13)$$

By the same analysis one obtains that the azimuthal electric field component vanishes.

The above equations relate the radial and poloidal electric field components to the gradient of magnetic flux and the angular velocity of the fluid into which is anchored a magnetic field. As a result, a notion of "velocity of magnetic field lines" is introduced. This is a poor choice of words, because it gives the impression of something physical, and is motivated by the fact that it is a constant along magnetic field lines[3]. In order to recover physical significance, it should be replaced by "velocity of the fluid frame in which the electric field vanishes". Alternatively, one could say that it can be "interpreted as an electromagnetic angular velocity"[8]. This induced electric field produces the emf that in this part of the circuit allows for the dissipation of energy (e.g. ref. [3]). More rigorously, in a force-free magnetosphere, the flow of energy from the black hole to infinity requires setting the covariant divergence of

$$\varepsilon_a G^{ab} \qquad (14)$$

to zero

$$\nabla_a \varepsilon_b G^{ab} = 0 \qquad (15)$$

where G is the electromagnetic energy-momentum tensor and ε solves the Killing equation to produce the conserved quantity in equation 14. The point is that the electric field components derived above appear in equation 15 which is then evaluated at the horizon. The point of equation (5) is to motivate the importance of those electric field components.

In the next section we show how these induced electric field components change when we implement a more realistic Ohm's law.

### 2.2 - Resistive MHD

A more realistic and familiar form of Ohm's law in standard vector notation is

$$\mathbf{J} = \sigma(\mathbf{E} + \mathbf{v} \times \mathbf{B}) \qquad (16)$$

where **J** is the current density vector and σ is the scalar conductivity. In covariant component form this is

$$J^a = \sigma F^{ab} U_b \qquad (17)$$

Using the same strategy of Section 2.1, we obtain Boyer-Lindquist electric field components, by evaluating

$$F_{rb}U^b = J_r/\sigma, \quad F_{\theta b}U^b = J_\theta/\sigma, \quad \text{and} \quad F_{\phi b}U^b = J_\phi/\sigma. \qquad (18)$$

Clearly, the induced electric fields are related to additional terms. The usual procedure is to start with the fields and obtain the current. But our goal is to think of using equation (17) to constrain or determine the electric fields. But what determines the electric and magnetic fields? The charge and current distributions, of course. In a numerical simulation, one fixes the conductivity and assumes an initial electromagnetic field. Then the simulation proceeds to update the fields which remain finite over time. The finite fields and their relation to finite current densities is captured by equation (17) and can be evaluated near the black hole. But there is no physics that can be identified that will short out the electric fields. What you bring to the black hole will determine the conditions near there despite the highly curved spacetime region. And the plasma that is approaching the horizon, is characterized by electric fields that are determined by the accreting physics and not by the black hole.

Apart from the above considerations, we wish to point out that although equation (17) may be better than equation (1) in terms of being more physical, it suffers from the fact that it violates special relativity[9]. In fact, equation (17) implies an instantaneous current in response to the fields. In the next section, we will further generalize Ohm's law as we attempt to make our physics even more realistic. In that context, we hope it will become clearer that any constraints on the electric field near the black hole will come from the complicated and non-intuitive physics of the fluid as governed by the generalized Ohm's law and not by ad-hoc considerations about the behavior of curved spacetime.

### 2.3 – *The generalized Ohm's law*

Because of the absence of any time dependence in the current in $J^a = \sigma F^{ab} U_b$, it must violate special relativistic causality. This has been long recognized [9,10,11]. The time dependence of the current appears in the generalized Ohm's law, which, for the reasons given above, we write in standard vector notation for a gas with electrons and ions as

$$(m_+ m_e c^2 / Z\rho e^2)\, \partial \mathbf{j}/\partial t \; + \mathbf{j}/\sigma = \mathbf{E} + \mathbf{v} \times \mathbf{B} \; + (c/eZ\rho)[m_+ \nabla p_e - m_e \nabla p_+ - (\, m_+ - Zm_e)\mathbf{j} \times \mathbf{B}]. \quad (19)$$

Above, $m_+$ is the mass of the ions, $m_e$ the mass of the electrons, c the speed of light, Z the atomic number, $\rho$ the mass density, $e$ the electric charge, and p the pressure of each particle species. Equation 19 is itself a simplification of a more general dynamical equation for the motion of charged particle species. Among the simplifications are (1) linearity, (2) a scalar pressure, (3) electrical neutrality. The last condition, in particular, would add terms to equation 19 by way of

$$n_+ Z - n_e \neq 0.$$

We emphasize that implementing the generalized Ohm's law is not simply a matter of adding one equation but a complex network of equations that constrain the quantities in equation 19. In attempting to understand the nature of the electric fields, Meier[9] points out that any number of phenomena may produce charge separation in the context of the generalized Ohm's law that are both causal and covariant. It is worth emphasizing that the state-of-the-art in simulations of accretion around black holes is still wedded either to equation (1) or at most to equation (17). The reasons for this is that turbulence allegedly makes much of the microphysics of equation 19 irrelevant. There are reasons to be skeptical about this especially close to the black hole horizon where magnetic fields are strong.

Despite the current limit in scope, recent particle-in-cell (PIC) simulations have begun to tease out the behavior of charged particles near the horizon and find that the Blandford-Znajek conditions are valid[12]. Komissarov[2] also addresses the King & Pringle critique of the Blandford-Znajek process in light of the work of ref. [12], concluding that although accumulation of electric charge onto the rotating black hole might affect dynamics in the magnetosphere, it cannot shut down the Blandford-Znajek process, but requires further investigation. He also emphasizes that the inner boundary conditions in PIC simulations (e.g., refs. [13, 12]) are accurate. From ref. [11] one can estimate that the resolution needed just for non-ideal (i.e. resistive) MHD effects, is 10 orders of magnitude smaller in length than the Schwarzchild radius, which to use a specific source as example, is $10^{15}$ cm for M87. Similarly, characteristic resistive MHD timescales are 10 orders of magnitude smaller than ideal MHD. The transit time for M87 is order $10^5$ seconds. The development of PIC simulations is of fundamental importance, especially because a full Ohm's law approach has not yet led to any tangible results. In short, we are quite removed from exploring the physics of the full Ohm's law that is needed to rigorously determine the structure of the electric field near the black hole. Until then, any discussion that attempts to invalidate the Blandford-Znajek effect, is moot.

As King & Pringle[1] point out, however, jets are observed in sources that do not have black holes. Hence, the energy for jets need not come from black hole rotational energy as they do in general relativistic MHD simulations of black holes. This does not mean that the rotational energy of the object in question is not responsible. If the rotational energy of the rotating object is solely responsible for jets, one expects a correlation between the angular velocity of the object and the jet power. Additionally, the accretion disk carries gravitational potential energy and can be tapped via the Blandford-Payne mechanism[14] by way of magnetic fields to produce a jet outflow. Evidence for an alternative source of energy is strongest in black hole X-ray binaries in so-called hard states, where no correlation is found between the black hole spin and the jet power [15].

3. **Summary and conclusions**

In this work we have shown that to assess the true electrical properties near rotating black holes in both analytic and numerical work, one must implement more realistic physics. From reviews of the state-of-the-art one acquires the impression that a physically well-motivated Ohm's law is not necessary (e.g. ref. [16]) and the reason for this is the ubiquitous belief that turbulence tends to make discussion of microphysics irrelevant. It is also important to emphasize that resolving the scales on which plasma effects make their appearance, is extremely computationally challenging. But while appealing to turbulent effects may be reasonable in much of a black hole magnetosphere, it is unlikely to be true in the region near the black hole where the intensity of gravity and the strength of the magnetic field are both large. If this is correct, we are left with an unresolved physical condition near black holes and no first-principles approach can resolve this.

**Acknowledgment**

We thank multiple anonymous referees for comments that improved the impact of this work. C.B.S. is supported by the National Natural Science Foundation of China (Grant No. 12073021).